\newcommand{\Real}{\mathbb{R}}
\newcommand{\Gel}{\mathbb{G}}
\newcommand{\Gbm}{\mbox{\bfseries\sffamily{G}}}
\newcommand{\Rbm}{\mbox{\bfseries\sffamily{R}}}
\newcommand{\Ratio}{\mathbb{Z}}
\newcommand{\caQ}{\mathcal Q}
\newcommand{\caE}{\mathcal E}
\newcommand{\caG}{\mathcal G}
\newcommand{\caD}{\mathcal D}
\newcommand{\caS}{\mathcal S}
\newcommand{\caM}{\mathcal M}

\def\dive{\mathop{\rm div\,}}
\def\diag{\mathop{\rm diag}}
\def\span{\mathop{\rm span}}
\newcommand{\dif}{\mathrm{d}}

\documentclass[10pt]{article}
\usepackage{amsfonts}
\usepackage{graphicx}

\begin{document}

\title{Existence, decay time and light yield for a reaction-diffusion-drift equation in the continuum physics of scintillators}

\author{Fabrizio Dav\'{\i}\footnote{E-mail: davi@univpm.it}\\
\em DICEA, Universit\'a Politecnica delle Marche, Ancona, Italy,
}

\date{}

\maketitle

\abstract{A scintillator is a material which converts incoming ionizing energy into visible light. This conversion process, which is a strongly nonlinear one, can be described by a Reaction-Diffusion-Drift equation we obtain from a model of continua with microstructure endowed with a suitable thermodynamics. For such an equation it can be show the global existence of renormalizable and weak solutions, and the solutions exponential decay estimates can be given; moreover we give also a mathematical definition for the light yield which is a measure of scintillation efficiency.}

\bigskip
\noindent
{\bf Keywords:} Reaction-Diffusion-Drift models, Scintillators, Entropy methods, Exponential rate of convergence.

\noindent
{\bf AMS Classification:} 35K57, 35B40, 35B45.

\section{Introduction}

A scintillator crystal is a "wavelength shifter" which converts energy, typically $\gamma-$rays, into photons in the frequency range of visible light. For this reason scintillators crystals are used in high-energy physics and in medical and security applications \cite{1}. The physics of scintillation is quite complex but it can be conveniently 
divided in three major phenomena which correspond to three different time and space scales: (i) the incoming energy generates an ionized region of few nanometers populated by charged energy carriers, a scale which we call the \emph{Microscopic} dealing with the creation of excitation carriers in the ionized region; (ii) these energy carriers generates other energy carriers within a greater region: when these carriers recombine a part of them generate photons. Such a phenomena evolves at a \emph{Mesoscopic} scale. (iii) the light rays propagates within the crystal, a phenomena which happens at a \emph{Macroscopic} scale.

In \cite{davi} we obtained, by means of a continuum with microstructure theory and a suitable thermodynamics, a model which describes the phenomena at the mesoscopic scale in terms of a Reaction-Diffusion Drift equation for the energy carriers descriptors, with Neumann-type boundary conditions, coupled with the heat and electrostatic equations. 

Such equation and its associated variational formulation are the starting point for the correct mathematical description of the two most important physical parameters which characterize a scintillator crystal: the \emph{Light Yield} $Y$, which is the ratio between the collected light energy and the energy of the incoming ionizing radiation (and which is indeed a measure of the scintillator efficiency) and the \emph{Scintillation Decay time} $\tau$ which is the time required for scintillation emission to decrease to $e^{-1}$ of its maximum and is a measure of the scintillator resolution.

Here first we show the main results obtained into \cite{davi} and then, by adapting the results of \cite{CHJU17} to the present formulation, we proof global existence of renormalized and weak solutions. Then by following the approach and the ideas of \cite{FK18} (\emph{vid. also\/} \cite{DEFE06}) we give an estimate for the decay time. Finally we propose a suitable definition for the light yield based on the decay time estimate.

\section{The evolution equation for scintillators}

\subsection{The excitation carrier density vector}

In order to define the basic state variable for our problem, we have to deal briefly with the scintillation phenomena at the microscopic scale and to the features of it which appears appears at the mesoscopic scale: we give here only the main ideas, the details which can be found in \cite{davi}. Basically, the incoming energy $E^{*}$ which hits the crystal at a point $x$ generates a great number of excitation carriers within a cylindrical track of radius $r$ and energy-dependent length $L=L(E^{*})$ from $x$: on this track we define an excitation density \cite{JA07}.  

In \cite{davi} we show that the relevant descriptor of the microscopic phenomena is the excitation density times the area of the cross-section of the cylindrical track and then, by "zooming-out" to the mesoscopic volume $\Omega$ centered on $x$ we get, by means of renormalization techniques,  the mesoscopic descriptor in $\Omega$:
\begin{equation}\label{N1}
N=\frac{1}{\pi r^{2}L(E^{*})}\frac{E^{*}}{E_{exc}}> 0\,,
\end{equation}
where $E_{exc}$ is the \emph{excitation energy}, which depends on the specific scintillator crystal.

Since the excitation carriers may exhibits different physical behaviour, \emph{i.e.} can recombine into photons or other kind of excitation carriers rather then annihilate themeselves in different ways, then $N$ can be decomposed into the sum of $k$ different kind of excitation carriers, the value of $k$ depending on how much we want a more detailed description of phenomena (for instance $k=2$ in \cite{LGW11} whereas $k\geq 11$ in \cite{VA08}):
\begin{equation}\label{N2}
N=\sum_{i=1}^{k}n_{i}\,,\quad n_{i}> 0\,.
\end{equation}
We find useful to introduce an \emph{excitation carrier density} $k$-dimensional vector as the basic state variable for our theory:
\begin{equation}
n\equiv(n_{1}\,,n_{2}\,,\ldots\,,n_{k})\,,\quad (x\,,t)\mapsto n_{j}(x\,,t)>0\,,\quad j=1\,,\ldots k\,;
\end{equation}
where $(x\,,t)\in\Omega\times [\mbox{0}\,,\tau)$ with $\Omega$ a mesoscopic control volume and $n\in\caM\equiv(0\,,\infty)^{k}$.

Let $q^{*}$ be the charge density associated to the incoming energy $E^{*}$, then such a charge and the excitation carriers generate a \emph{local electric field} $(x\,,t)\mapsto\varphi(x\,,t)$:
\begin{equation}\label{LEF}
-\epsilon_{o}\Delta\varphi=q^{*}+e\,z\cdot n\,,\quad\mbox{in }\Omega\times [\mbox{0}\,,\tau)\,,
\end{equation}
with Neumann-type boundary conditions on $\partial \Omega\times [\mbox{0}\,,\tau)$: here $e$ is the elementary electron charge, $\epsilon_{o}>0$ is the vacuum permittivity and $z\in\Ratio^{k}$ is the \emph{charge vector}.

\subsection{The reaction-diffusion-drift equation}

To model the recombination of excitation carriers within $\Omega$, in \cite{davi} we wrote the equation of electric current balance in terms of the theory of continua with microstructure (\emph{vid. e.g.\/} \cite{CA00}) endowed with a suitable thermodynamic and appropriate constitutive hypotheses. In particular we assumed a Gibbs free-energy
\begin{equation}
\psi=\varepsilon-\theta\eta
\end{equation}
with internal energy $\varepsilon$ and entropy $\eta$ given respectively by:
\begin{equation}
\varepsilon(n\,,\theta)=e\varphi\,z\cdot n+u(\theta)\,,
\end{equation}
and
\begin{equation}\label{entropia}
\eta(n\,,\theta)=-k_{B}\sum_{i=1}^{k}n_{i}(\log C_{i}n_{i}-1)+\lambda\log\theta\,,
\end{equation}
where $k_{B}$ is the Boltzmann constant, $\lambda>0$ the latent heat and $C_{i}>0$ are normalizing constant. In the model we obtained, the only interaction with the macroscopic scale is the \emph{absolute temperature} $\theta=\theta(x\,,t)>0$: however a wider range of macroscopic variables, like mechanical strain, crystal defects and electromagnetic fields (as in \cite{XB08}, \emph{e.g.}) will be dealt with in a forthcoming paper \cite{DA172}.

From the dissipation inequality, constitutive assumptions and the balance laws for a continuum with microstructure we arrive at an equation which describes the generation and recombination of excitation carriers:
\begin{equation}\label{RDD}
\mbox{div}(D\nabla n+MNz\otimes\nabla\varphi)-K(n)n=\dot{n}\,,\quad\mbox{in }\Omega\times [\mbox{0}\,,\tau)\,,
\end{equation}
which is a rection-diffusion-drift equation with Neumann boundary condition on $\partial \Omega\times [\mbox{0}\,,\tau)$ and initial conditions $n(x\,,0)=n_{0}(x)$ which depend, by means of (\ref{N1}) and (\ref{N2}), on the incoming energy $E^{*}$ at $x$. 

The various terms in (\ref{RDD}) represent:
\begin{itemize}
\item $N$ is the $k\times k$ matrix $N=\mbox{diag}(n_{1}\,,n_{2}\,,\ldots\,,n_{k})$;
\item $M=M(\theta)$ is the $k\times k$ symmetric and positive-definite \emph{Mobility matrix};
\item $D=(k_{B}\theta/e) M$, is the $k\times k$ \emph{Diffusivity matrix};
\item $\varphi$ is the \emph{local electric field} solution of (\ref{LEF});
\item $K=K(n\,,\theta)$ is a non-linear function of $n$ which describes the recombination process.
\end{itemize}

Equation (\ref{RDD}) is coupled with (\ref{LEF}) and with the heat equation (with an electrostatic source term) 
\begin{equation}
\dot{\theta}=\dive\mathbf{C}\nabla\theta-ez\cdot\dot{n}\,,\quad\mbox{in }\Omega\times [\mbox{0}\,,\tau)\,,
\end{equation}
with Neumann boundary conditions on $\partial \Omega\times [\mbox{0}\,,\tau)$; here $\mathbf{C}$ is the positive-definite crystal \emph{Conductivity tensor}.

Equation (\ref{RDD}) generalizes the two most important phenomenological models for scintillation, namely the \emph{Kinetic} (\emph{vid. e.g.} \cite{BM09}) and the \emph{Diffusion} models \cite{BM12}: they are the same equations postulated into \cite{VA08} and used into \cite{LGW17} to perform numerical analysis of solutions; they are also are identical (apart for the reaction term $K(n)n$) to the equations for the semiconductors obtained, by starting from a different approach, into \cite{AGH02}, \cite{MIE11} and \cite{MIE15}.

Thermodynamics allows to write the \emph{Dissipation} associated to equation (\ref{RDD}):
\begin{equation}
\caD=2\Psi(n\,,\mu\,,\theta)\,,
\end{equation}
where the \emph{Conjugate dissipation functional} is defined as
\begin{equation}\label{dissi}
\Psi(n\,,\mu\,,\theta)=\frac{1}{2}\int_{\Omega}S(n\,,\theta)[\nabla\mu]\cdot\nabla\mu+H(n\,,\theta)\mu\cdot\mu>0\,;
\end{equation}
here $\mu$ is the \emph{Scintillation potential} (indeed the equivalent for scintillators of the electrochemical potential in semiconductors)
\begin{equation}\label{scintpot}
\mu=\frac{\partial\psi}{\partial n}=e\varphi z+k_{B}\theta\log(n^{*})\,,\quad\log(n^{*})\equiv(\log C_{1}n_{1}\,,\ldots\log C_{k}n_{k})\,,
\end{equation}
and where the positive-definite $k\times k$ matrices $S$ and $H$ are given by $S(n\,,\theta)=e^{-1}M(\theta)N(n)$ and $H(n\,,\theta)\mu=K(n\,,\theta)n$.

We notice that relation (\ref{scintpot}) can be inverted to obtain
\begin{equation}\label{npotential}
n=\Lambda\,c\,,
\end{equation}
with $\Lambda=(e^{\frac{1}{k_{B}\theta}(\mu_{1}-ez_{1}\varphi)}\,,\ldots\,,e^{\frac{1}{k_{B}\theta}(\mu_{k}-ez_{k}\varphi)})$ and  $c\equiv(c_{1}\,,\ldots\,,c_{k})$ with $c_{i}=C_{i}^{-1}$.

By means of (\ref{dissi}), equation (\ref{RDD}) can be put in the equivalent gradient flow formulation, namely:
\begin{equation}\label{gradientflow}
\dot{n}=-D\Psi(n\,,\mu\,,\theta)\,,
\end{equation}
where $D\Psi$ denotes the Frechet derivative of the dissipation $\Psi$; notice that (\ref{RDD}) can be expressed in terms of the scintillation potential as:
\begin{equation}\label{RDD1}
\dive S[\nabla\mu]-H\mu=\dot{n}\,,
\end{equation}
a form we shall make use of in the sequel.

\section{Existence, Decay time estimates and Light yield}

Trought this section we shall deal with isothermal scintillators, in such a way that the fixed temperature $\theta=\theta_{o}$ appears only as a parameter in the constitutive terms $M$ and $K$ and the problem is described by equations (\ref{LEF}) and (\ref{RDD}) only: moreover we shall assume that the domain can be rescaled by a characteristic length $l^{*}$ and w.l.o.g. we set that the adimensional parameter
\begin{equation}
\beta=\frac{k_{B}\theta_{o}}{el^{*}}=1\,,
\end{equation}
in such a way that $D=M$. With a slight abuse of notation we shall still denote $\Omega$ the rescaled domain.

\subsection{Global existence}

The problem of finding existence, asymptotic estimates and qualitative bounds for the solutions for the reaction-diffusion drift equations like (\ref{RDD}) coupled with (\ref{LEF}) has received a strong attention in the recent years, \emph{vid. e.g.\/} \cite{CHJU17}-\cite{DEFE06}, \cite{GA94}-\cite{FI17}: to this regard it is important to remark that most of them deal with semiconductors or chemical reactions which differ from scintillators by the reaction term $K(n)n$.

In \cite{CHJU17} however, a global existence for (\ref{RDD}) with Neumann-type boundary conditions in terms of \emph{renormalized solutions} was obtained for a general reaction term $K(n)n$. According to the definition given into \cite{FI17}, renormalized solutions $n$  to the reaction-diffusion-drift
equation (\ref{RDD})  are defined by the condition that for all functions $\xi:\caM\rightarrow\Real$  with compactly
supported derivative $\nabla_{n}\xi$ , the function $\xi(n)$ must satisfy the equation derived from (\ref{RDD}) by a formal application of the chain rule in a weak sense. 

More precisely, according to \cite{CHJU17} and \cite{FI17}, $n\equiv(n_{1}\,,n_{2}\,,\ldots\,,n_{k})$  is a \emph{renormalized solutions} for (\ref{RDD}) if $\forall \tau>0$, $n_{i}\in L^{2}([0\,,\tau); H^{1}(\Omega))$ and for any $\xi\in C^{\infty}(\caM)$ satisfying $\nabla_{n}\xi\in C_{0}^{\infty}(\caM; \Real^{k})$ and $\psi\in C_{0}^{\infty}(\bar{\Omega}\times [0\,,\tau))$ it holds:
\begin{eqnarray}
&&\int_{0}^{\tau}\int_{\Omega}\xi(n)\dot{\psi}+\int_{\Omega}\xi(n)\psi\big{|}_{0}^{\tau}=\\
&+&\int_{0}^{\tau}\int_{\Omega}(\nabla\nabla\xi\cdot S[\nabla\mu]\otimes\nabla n)\psi+\int_{0}^{\tau}\int_{\Omega}S[\nabla\mu]\nabla\xi\cdot\nabla\psi\nonumber\\
&+&\int_{0}^{\tau}\int_{\Omega}(H\mu\cdot\nabla\xi)\psi,.\nonumber
\end{eqnarray}

Let $\caE$ be the \emph{total scintillation entropy} on the control volume $\Omega$:
\begin{equation}\label{entropytotal}
\caE(n)=-\int_{\Omega}\sum_{i=1}^{k}n_{i}(\log C_{i}n_{i}-1)\,,
\end{equation}
then the main result of \cite{CHJU17} rephrased in terms of (\ref{RDD}) states that, provided the following hypotheses hold:
\begin{itemize}
\item [(H1)] Drift term: $\nabla\varphi\in L^{\infty}([0\,,\tau)\,; L^{\infty}(\Omega))$;
\item [(H2)] Reaction term: $K(n)n\in C_{0}([0\,,\tau)^{k}; \caM)$;
\item [(H3)] Initial data: $n_{o}\equiv(n^{0}_{1}\,,n^{0}_{2}\,,\ldots\,,n^{0}_{k})$ is measurable, $n^{0}_{i}> 0$ in $\Omega$, $i=1,2,\ldots k$ and
\[
\caE(n_{o})<+\infty\,;
\]
\item [(H4)] There exist numbers $\pi_{i}>0$ and $\lambda_{i}\in\Real$, $i=1,2,\ldots,k$ such that for all $n\equiv(n_{1}\,,n_{2}\,,\ldots\,,n_{k})\in\caM$, the following inequality holds:
\[
\sum_{i=1}^{k}\pi_{i}(K(n)n)_{i}(C_{i}\log n_{i}+\lambda_{i})\leq 0\,;
\]
\item [(H5)] The mobility matrix $M$ is symmetric and positive-definite;
\end{itemize}
then equation (\ref{RDD}) admits a renormalized solution $n\equiv(n_{1}\,,n_{2}\,,\ldots\,,n_{k})$ satisfying $n_{i}>0$ in $\Omega$, $i=1,2,\ldots k$ and\begin{equation}
\caE(n)=<+\infty\,;\quad\forall t>0\,.
\end{equation}

As pointed out in \cite{FI17}, moreover, any renormalized solution for which $K(n)n=H\mu\in L^{1}([0\,,\infty)^{k}; \caM)$ is also a weak solution of (\ref{gradientflow}) in the sense that, for any $v\equiv(v_{1}\,,\ldots v_{k})\in C^{\infty}([0\,,\tau)^{k}\,,\caM)$:
\begin{equation}
\int_{\Omega}(v\cdot n)\big{|}_{0}^{\tau}-\int_{0}^{\tau}\int_{\Omega}n\cdot\dot{v}=-\int_{0}^{\tau}\int_{\Omega}S[\nabla\mu]\cdot\nabla v+H\mu\cdot v\,;
\end{equation}
as far as we know we can instead  say nothing about the global existence in time of smooth solutions.

\subsection{Decay time}

The available experimental data (\emph{vid. e.g.} the recent analysis in \cite{SWI14}) and the numerical solution of phenomenological models as in \cite{LGW17}, show that the excitation carriers decay exponentially to an asymptotic homogeneous value $n_{\infty}$, namely:
\begin{equation}
\|n(\cdot\,,t)-n_{\infty}(\cdot)\|=A_{f}\exp(-t/\tau_{f})+A_{s}\exp(-t/\tau_{s})\,,
\end{equation}
where the indeces $f$ and $s$ denotes the so-called \emph{fast} and \emph{slow} components of the excitation, respectively. Accordingly, since by definition the \emph{Decay time} is the time required for scintillation emission to decrease to $e^{-1}$ of its maximum, then we get a Fast Decay Time $\tau_{f}$ and a Slow Decay Time $\tau_{s}$.

In many cases one of the components is negligible and the decay obeys a simple exponential law, which can be also used to describe an average decay time. To this regard, in \cite{FK18}, \cite{DEFE06}, \cite{DEFE08}, \cite{FETA17} and \cite{FK18a}, an explicit estimate of the asymptotic convergence was obtained for the cases of chemical reactions and semiconductors. In particular in \cite{FK18} the Rosbroeck model with Shockley-Read-Hall potential for semiconductors was studied. In the following we shall se how the same approach and ideas of \cite{FK18} can be extended to the case of scintillators in order to obtain an explicit estimate for the decay time. 

Scintillation depends on the evolution of charge carriers: accordingly we must require that trough the whole process the electric charge is conserved. Accordingly, let $\caQ\in\Real$ be the \emph{total electric charge} 
\begin{equation}
\caQ(n)=\caQ^{*}+\int_{\Omega}ez\cdot n\,,\quad \caQ^{*}=\int_{\Omega}q^{*}\,,
\end{equation}
then by (\ref{LEF}) with Neumann boundary condition we must have:
\begin{equation}\label{cons}
\caQ^{*}+\int_{\Omega}ez\cdot n=0\,,\quad \forall t\in[0\,,\tau)\,.
\end{equation}
We remark that (\ref{cons}) is the necessary condition to have an unique weak solution $\varphi\in H^{1}(\Omega)$ to equation (\ref{LEF}) with Neumann boundary conditions and such that $\overline{\varphi}=0$, where 
\[
\overline{f}=\frac{1}{\mbox{vol}\,\Omega}\int_{\Omega}f\,,
\]
denotes the mean value on $\Omega$.

Moreover (\ref{cons}) leads to the conservation law
\begin{equation}\label{chargecons}
\frac{\dif}{\dif t}\caQ=\int_{\Omega}ez\cdot\dot{n}=0\,, \quad\forall t\in[0\,,\tau)\,,
\end{equation}
and from (\ref{chargecons})$_{1}$ and (\ref{RDD}) with Neumann-type boundary conditions:
\begin{equation}\label{chargecons1}
\int_{\Omega}K(n)n\cdot z=0\,,\quad\forall t\in[0\,,\tau)\,.
\end{equation}

It is important to remark that the total charge $\caQ$ depends on the type of ionizing radiation which hits the scintillator: indeed for $\gamma-$ and $X-$rays we have $\caQ^{*}=0$, whereas for $\alpha-$rays it is $\caQ^{*}>0$ and $\caQ^{*}<0$ for $\beta-$rays.

Let $n_{\infty}(x)$ and $\varphi_{\infty}(x)$ be the stationary solution(s) of (\ref{RDD}) and (\ref{LEF}), \emph{i.e.\/} with $\dot{n}=0$ (\emph{cf.\/} \cite{WMZ08}); it is easy to see from (\ref{RDD1}) that for the stationary solutions the scintillation potential vanishes \emph{i.e.\/}
\begin{equation}\label{munull}
\mu_{\infty}=0\,.
\end{equation}
Accordingly, from (\ref{npotential}), by (\ref{scintpot})  we have:
\begin{equation}
n_{\infty}=F_{\infty}c\,,
\end{equation}
where $F_{\infty}=\diag(e^{-ez^{*}_{1}\varphi_{\infty}}\,,\ldots\,,e^{-ez^{*}_{k}\varphi_{\infty}})$, $z^{*}_{j}=ez_{j}/k_{B}\theta_{o}$ and with $\varphi_{\infty}$ the unique solution of the Neumann-type problem \cite{WMZ08}
\begin{equation}\label{LEF1}
-\epsilon_{o}\Delta\varphi_{\infty}=q^{*}+e\,z\cdot F_{\infty}c\,,\quad\mbox{in }\Omega\,,\quad\overline{\varphi}_{\infty}=0\,,
\end{equation}
provided
\begin{equation}\label{cons1}
-\caQ^{*}=ez\cdot\int_{\Omega}n_{\infty}=ez\cdot\overline{F}_{\infty}c\,,
\end{equation}
holds; we remark that condition (\ref{munull}) trivially verifies both (\ref{chargecons1}) and:
\begin{equation}
z\cdot\overline{K(n_{\infty})n_{\infty}}=0\,.
\end{equation}
We notice that, in the case of $\gamma-$rays we have $\caQ^{*}=0$ and condition (\ref{cons1}) implies that, for  $\caS\equiv\span\{(\overline{F}_{\infty})^{T}z\}$, then $c\in\caS^{\perp}$.

In order to grant uniqueness for $c$ and hence for $(n_{\infty}\,,\varphi_{\infty})$, we need additional hypotheses on the reaction term $K(n)n$, as it was done  in \cite{FK18} for the case of Rosbroeck semiconductors with $k=2$, where for the reaction term was assumed a Shockley-Read-Hall potential (\emph{vid. also\/} \cite{FK18a}). Here we simply assume as a constitutive prescription that the reaction term $K(n)$ is such that $c$ is unique and there exists two positive constants $K_{1,2}$ such that:
\begin{equation}\label{bound0}
K_{1}\leq\|K(n)\|_{L^{\infty}(\Omega)}\leq K_{1}+K_{2}\|n_{o}\|_{L^{\infty}(\Omega)}^{m}=K_{\infty}\,,\quad m>1\,;
\end{equation}
then we may assume that the following bounds for $c$ and $n_{\infty}$ hold (\emph{cf.\/}\cite{FK18}):
\begin{equation}\label{bounds}	
\|c\|\leq K_{\infty}e^{\Phi_{\infty}}(1+|Q^{*}|)\,,\quad\|n_{\infty}\|\leq K_{\infty}e^{2\Phi_{\infty}}(1+|Q^{*}|)\,,
\end{equation}
with $\Phi_{\infty}=e\|z\varphi_{\infty}\|_{L^{\infty}(\Omega)}$.

The total Gibbs free-energy for a scintillator is given by
\begin{equation}
\caG(n\,,\varphi(n))=\int_{\Omega}\psi(n\,,\varphi(n))=\int_{\Omega}\varepsilon(n\,,\varphi(n))-\theta_{o}\eta(n\,,\varphi(n))\,;
\end{equation}
then following \cite{FK18} (\emph{see also\/} \cite{FI17}), we define the \emph{Relative Gibbs free-energy } (which in \cite{FK18} is referred as "relative entropy") as:
\begin{equation}
\caG(u|v)=\caG(u)-\caG(v)-D\caG(v)(u-v)\,.
\end{equation}
By an explicit calculation we obtain:
\begin{equation}
\caG(n|n_{\infty})=\int_{\Omega}\sum_{i=1}^{k}n_{i}\log(\frac{n_{i}}{n^{\infty}_{i}})+(n^{\infty}_{i}-n_{i})+\frac{1}{2}\varepsilon_{o}\|\nabla\varphi-\nabla\varphi_{\infty}\|^{2}\,,
\end{equation}
and then, by an easy calculation, it can be shown that the \emph{Dissipation} $\caD$ is given by \cite{davi}:
\begin{equation}\label{dissipation}
\caD=-\frac{\dif}{\dif t}\caG=2\Psi(n\,,\mu)>0\,.
\end{equation}

We follow \cite{FK18} and by starting from (\ref{dissipation}), by means of a repeated use of Csisz\'ar-Kullback-Pinsker type inequalities we may arrive, provided (\ref{bounds}) hold, to the following estimates for the case $k=2$:
\begin{eqnarray}\label{estimate0}
&\caD(n\,,\varphi(n))\geq C_{1}\caG(n\,,\varphi(n))\,,\nonumber\\
\\
&\|n-n_{\infty}\|^{2}_{L^{1}(\Omega)}+\|\varphi-\varphi_{\infty}\|^{2}_{H^{1}(\Omega)}\leq C_{2}\caG(n_{o}\,,\varphi_{o})e^{-C_{1}t}\,,\nonumber
\end{eqnarray}
with $\varphi_{o}$ the unique solution of 
\begin{equation}\label{LEF0}
-\epsilon_{o}\Delta\varphi_{o}=q^{*}+e\,z\cdot n_{o}\,,\quad\mbox{in }\Omega\,,\quad\overline{\varphi}_{o}=0\,,
\end{equation}
with Neumann-type boundary conditions and where the parameters $C_{1,2}$ have the explicit expression \cite{FK18}:
\begin{eqnarray}\label{parameter}
C_{1}^{-1}&=&\frac{1}{2}K_{\infty}e^{2\Phi_{\infty}}(1+|Q^{*}|)\max\{\frac{\epsilon_{o}}{M^{*}}K_{\infty}e^{2\Phi_{\infty}}(1+|Q^{*}|)\,,\frac{1}{K_{1}}\}\cdot\nonumber\\
&&\cdot(1+\frac{L(\Omega)}{\epsilon_{o}}K_{\infty}e^{2\Phi_{\infty}}(1+|Q^{*}|))\,,\\
C_{2}&=&(3K_{\infty}e^{2\Phi_{\infty}}(1+|Q^{*}|)+\frac{1}{2}\caG(n_{o}\,,\varphi_{o})+\frac{2}{\varepsilon_{o}}(1+L(\Omega))\,,\nonumber
\end{eqnarray}
where $L(\Omega)$ is the Poincar\'e constant of $\Omega$ and $M^{*}$ is the smallest eigenvalue of $M$.

The expression for the decay time $\tau=C_{1}^{-1}$ depends, by (\ref{parameter})$_{1}$, in an explicit manner on the mobility $M$, the reaction term $K(n)$, the initial data $n_{o}$, the charge $Q^{*}$ and the scintillation volume $\Omega$. The extension to the case $k>2$ and to specific expression for $K(n)$ will be the object of further studies: however, as fare as we know, this is the first explicit estimate of the decay time in term of the problem physical (and measurable) parameters.

\subsection{Light yield}

In order to define the light yield we must be able to discriminate the recombinations of excitation carriers which converts into photons from those which exhibit "quenching", that is recombination without emission. To this regard in the most successful phenomenological model for scintillator, the "Kinetic model", borrowed from chemical reactions (\emph{vid. e.g.} \cite{1}, \cite{VA08}, \cite{BM09}),  the matrix $K(n)$ was  assumed as a quadratic function of $n$: 
\begin{equation}
K_{ij}(n)=R_{ij}+G_{ij}+E_{ij}+(\Rbm_{ijh}+\Gbm_{ijh})n_{h}+\Gel_{ijhm}n_{h}n_{m}\,,
\end{equation}
$i,j,h,m=1\,,2\,,\ldots k$, where the terms $R_{ij}$ and $\Rbm_{ijh}$ account for the linear and quadratic recombination, the terms $G_{ij}\,,\Gbm_{ijh}$ and $\Gel_{ijhm}$ accounts for the linear, quadratic and cubic (Auger) quenching whereas the exchange matrix $E_{ij}$ accounts for the excitation carriers which converts in other types. We remark that in this case bound (\ref{bound0}) hold for $m=2$ with $K_{1}\approx\|R+G+E\|$ and $K_{2}\approx\|\Rbm+\Gbm\|+\|\Gel\|$.

The most accepted definition of light yield in terms of the parameters of the phenomenological model is given  \emph{e.g\/} in \cite{BMSVW09}; let $n_{p}(x\,,t)$ be the solution of (\ref{RDD}) for $G_{ij}=0$, $\Gbm_{ijh}=0$ and $\Gel_{ijhm}=0$, \emph{i.e.\/} the solution which converts into visible light photons and let:
\begin{equation}
N_{p}(x\,,t)=\sum_{j=1}^{k}n_{j}^{p}(x\,,t)\,,\quad N_{o}(x)=\sum_{j=1}^{k}n_{j}^{o}(x)\,;
\end{equation}
then we define the \emph{Local light yield} $Y_{L}$ at a given point $\bar{x}$ as:
\begin{equation}
Y_{L}(\bar{x})=\frac{1}{\bar{\tau} N_{o}(\bar{x})}\int_{0}^{\bar{\tau}}N_{p}(\bar{x}\,,t)\dif t\,,
\end{equation}
where the characteristic time $\bar{\tau}$ is sometimes assumed as
\begin{equation}
\bar{\tau}^{-1}=\sup\{R_{ij}\}\,.
\end{equation}

A \emph{Global light yield} can be defined taking into account a characteristic volume, either the volume of the track (as in \cite{BM09}) or the scintillation volume $\Omega$ about $\bar{x}$:
\begin{equation}
Y(\Omega)=\int_{\Omega}\frac{1}{\bar{\tau} N_{o}(\bar{x})}\int_{0}^{\bar{\tau}}N_{p}(\bar{x}\,,t)\dif t\,.
\end{equation}

We propose here a different definition for the global light yield, based on the results of the previous section. Let
\begin{equation}
\bar{N}_{o}=\|n_{o}(x)-n_{\infty}(x)\|_{L^{1}(\Omega)}\,,\quad \bar{N}_{p}(t)=\|n_{p}(x\,,t)-n_{\infty}(x)\|_{L^{1}(\Omega)}\,,
\end{equation}
then the bound (\ref{estimate0}) holds and we may define an estimate for the global light yield:
\begin{equation}
Y(\Omega)=\frac{1}{\bar{\tau}\bar{N}_{o}}\int_{0}^{\bar{\tau}}\bar{N}_{p}(t)\dif t\leq \frac{1}{\bar{\tau}\bar{N}_{o}}\int_{0}^{\bar{\tau}}C_{2}\caG(n_{o}\,,\varphi_{o})e^{-C_{1}t}\dif t\,,
\end{equation}
to arrive at:
\begin{equation}
Y(\Omega)\leq\frac{C_{2}}{C_{1}}(1-e^{-C_{1}\bar{\tau}})\caG(n_{o}\,,\varphi_{o})\,,
\end{equation}
with $C_{1,2}$ evaluated for $G_{ij}=0$, $\Gbm_{ijh}=0$ and $\Gel_{ijhm}=0$. A further analysis of such  definition and the study of its relation with the classical one will be done in the future.

\section*{Acknowledgements}

\noindent The research leading to these results is within the scope of CERN R\&D Experiment 18 "Crystal Clear Collaboration" and has received funding
from the European Research Council under the COST action TD-1401 "FAST - Fast Advanced Scintillation Timing". The author wishes to thanks K. Fellner for pointing his attention on Ref. \cite{FK18} and \cite{FK18a}.

\end{document}